\documentclass[manuscript]{acmart}
\usepackage{mathtools,halloweenmath}
\AtBeginDocument{%
  \providecommand\BibTeX{{%
    \normalfont B\kern-0.5em{\scshape i\kern-0.25em b}\kern-0.8em\TeX}}}

\acmConference[CHI '22]{CHI '22:  Workshop on
Generative AI and HCI}{April, 2022}{}
\acmBooktitle{CHI '22: Workshop on
Generative AI and HCI,
  April, 2022}

\begin{document}

\title{Towards Co-Creative Generative Adversarial Networks for Fashion Designers}

\author{Imke Grabe}
\email{imgr@itu.dk}
\author{Jichen Zhu}
\email{jicz@itu.dk}
\affiliation{%
  \institution{IT University of Copenhagen}
  \streetaddress{Rued Langgaards Vej 7}
  \city{Copenhagen}
  \country{Denmark}
  \postcode{2300}
}




\begin{abstract}
Originating from the premise that Generative Adversarial Networks (GANs) enrich creative processes rather than diluting them, we describe an ongoing PhD project that proposes to study GANs in a co-creative context.
By asking 
\textit{How can GANs be applied in co-creation, and in doing so, how can they contribute to fashion design processes?} the project sets out to investigate co-creative GAN applications and further develop them for the specific application area of fashion design.
We do so by drawing on the field of mixed-initiative co-creation. 
Combined with the technical insight into GANs' functioning, we aim to
understand how their algorithmic properties translate into interactive interfaces for co-creation and propose new interactions.

\end{abstract}
\keywords{generative adversarial networks, mixed-initiative co-creation, human-AI collaboration}
\maketitle

\section{Introduction}

With new technological inventions occurring over time, such as the mechanical loom during the industrial revolution, new possibilities of creating cultural artifacts, such as clothing items, emerge.
In the age of industrialization 4.0, we now see how data-driven technologies increasingly find a place in the production cycle of clothing artifacts.\footnote{
\href{https://www.forbes.com/sites/brookerobertsislam/2021/01/27/zara-meets-netflix-the-fashion-house-where-ai-replaces-designers-eliminating-overstock/}{https://www.forbes.com/sites/brookerobertsislam/2021/01/27/zara-meets-netflix-the-fashion-house-where-ai-replaces-designers-eliminating-overstock/}}
With the recent advances in generative machine learning, human-made technology has gotten to a point where tools can \textit{imagine} complex outputs 
resembling the properties of real objects, such as art~\cite{elgammal_can_2017}, faces~\cite{karras_style-based_2019}, or clothing outfits~\cite{rostamzadeh_fashion-gen_2018}.
Latent variable models like Generative Adversarial Networks (GANs) learn to generate high-dimensional artifacts given a latent code as input~\cite{goodfellow_generative_2014}. Via its multiple neural layers, the generator network links the latent codes to output features resembling the training data.
The emerging entangled latent space encodes the learned semantics.
Due to the latent space's complexity, the connection between latent codes and output features is not traceable for the human eye.
While this complex, non-linear structure of encoded semantics impedes the control of generated designs, the emerging design space also makes GANs a novel tool offering new avenues for (co-)creation.

However, with new models for interaction come new challenges \cite{buschek_nine_2021}, primarily caused by the knowledge gap between machine learning models and their potential users.
Next to the ongoing research investigating the algorithmic properties of such models~\cite{wu_survey_2017,shamsolmoali_image_2021,roy_survey_2021,zhang_visual_2018}, a theoretical understanding of how to enfold their potential as creative collaborators in design processes is yet to be developed.
Originating from the premise that GANs enrich creative processes rather than diluting them, the ongoing PhD project presented here proposes to study GANs in a co-creative context.
By asking 
\textit{How can GANs be applied in co-creation, and in doing so, how can they contribute to fashion design processes?} the project sets out to investigate co-creative GAN applications and further develop them for the specific application area of fashion design.
We do so by drawing on the field of mixed-initiative co-creation. 
Combined with the technical insight into GANs' functioning, we aim to
understand how its algorithmic properties translate into interactive interfaces for co-creation and propose new interactions.

In the following, we provide an overview of related work to motivate the project, describe the work-in-progress, and conclude with reflections and open questions.

\section{Background}
Caused by the lack of interpretability underlying deep neural networks, 
designing human-AI interaction remains an open issue in the field of Human-Computer Interaction~\cite{yang_re-examining_2020}, in which GANs are often applied for providing variation~\cite{schrum_interactive_2020}.
Utilizing deep neural networks' black-box characteristics as a source of ``unpredictability''\footnote{\href{https://www.forbes.com/sites/brookerobertsislam/2020/09/21/why-fashion-needs-more-imagination-when-it-comes-to-using-artificial-intelligence/}{https://www.forbes.com/sites/brookerobertsislam/2020/09/21/why-fashion-needs-more-imagination-when-it-comes-to-using-artificial-intelligence/}} might add interesting aspects to design processes, 
focusing on the uncertainty in probabilistic machine learning.
\citeauthor{benjamin_machine_2021}~\cite{benjamin_machine_2021} approach machine learning uncertainty as a design material by proposing a phenomenological analysis of how machine learning models inferred from data affect our relation to the world.
How this applies to generative models, often applied for adding randomness to design processes, is yet to be explored. 
In the specific case of GANs, \citeauthor{hughes_generative_2021}~\cite{hughes_generative_2021} find variation as one of the main modes of operation when applied in design tasks, next to beautification.
As the first systematic analysis in the area, their survey categorizes what GANs add to design processes in different creative domains.
When it comes to interaction, 
GANs bring new challenges that require specific attention~\cite{buschek_nine_2021}.

To understand how GANs can be embedded into co-creation despite functioning as a course of randomness, one needs to turn towards the algorithmic properties behind interactive interfaces. 
Existing tools like GANLab\footnote{https://poloclub.github.io/ganlab/}
can to some extend reveal the system mechanics of a black-boxed model with two-dimensional data. However, it becomes difficult to visualize high-dimensional distributions like images,
let alone to control the space of possible output images.
To navigate the design space of latent variable models with more dimensions, several algorithmic proposals are of relevance to and have been tried out in interactive scenarios.
\citeauthor{karras_style-based_2019}~\cite{karras_style-based_2019} suggest the interpretation of the latent space entanglement by
analyzing how generated images respond to interpolations in the latent space aligned with human perception of change (measured as perceptual path length).
Applying interpolation in an interactive interface, humans can traverse between artifacts in latent space to explore intermediate versions~\cite{schrum_interactive_2020,gajdacz_creablender_2021}.
Finding hyperplanes that separate the latent vectors corresponding to output faces with and without certain attributes (measured as linear separability)~\cite{karras_style-based_2019} 
allows for the identification of attribute vectors~\cite{shen_interpreting_2020}. These vectors can then be used to manipulate the respective characteristic in semantic editing of e.g. faces~\cite{zaltron_cg-gan_2020,denton_image_2019}. 
By embedding GANs into an interactive genetic algortihm, 
\citeauthor{bontrager_deep_2018}~\cite{bontrager_deep_2018} present a paradigm for controlling the latent space of GANs with latent vector evolution~\cite{bontrager_deepmasterprints_2018}. By selecting images (phenotypes), the user can guide the next generation of artifacts, such as shoes, by indirectly evolving the corresponding latent codes (genotypes). In a similar process, \citeauthor{xin_object_2021}~\cite{xin_object_2021} apply conditional GANs to prompt the generation of more specific items, e.g. given the contour of a shoe.
Using the disentanglement of secondary latent space \cite{karras_style-based_2019} instead,
\citeauthor{tejeda_ocampo_appendix_2021}~\cite{tejeda_ocampo_appendix_2021} improve the model for the generation of images with more specific constraints due to more control over specific features.
These examples originate from investigating algorithmic possibilities and testing them in practice. However, the lack of a deeper understanding for the processes at play in co-creation with GANs as well as in considering designers' needs outlines the knowledge gap between machine learning engineers and designers~\cite{hughes_generative_2021}.

Born in the area of games, the research field of mixed-initiative co-creation aims to shed light on how human and computational agents create artifacts together. 
\citeauthor{spoto_library_2017}~\cite{spoto_library_2017} and \citeauthor{deterding_mixed-initiative_2017}~\cite{deterding_mixed-initiative_2017} propose a set of actions to map the interactions between the agents. Expanded into a framework for analyzing generative AI,
\citeauthor{muller_mixed_2020}~\cite{muller_mixed_2020} suggest to find interaction patterns when designing in generative design spaces. By using theoretical frameworks for understanding mixed-initiate co-creation, we aim to connect the algorithmic development of GANs with their role in the co-creation of artifacts with humans, ultimately allowing us to develop meaningful applications for our use cases in fashion design.

\section{Work-in-Progress and Reflections}

\begin{figure}
    \centering
    \includegraphics[width=15cm]{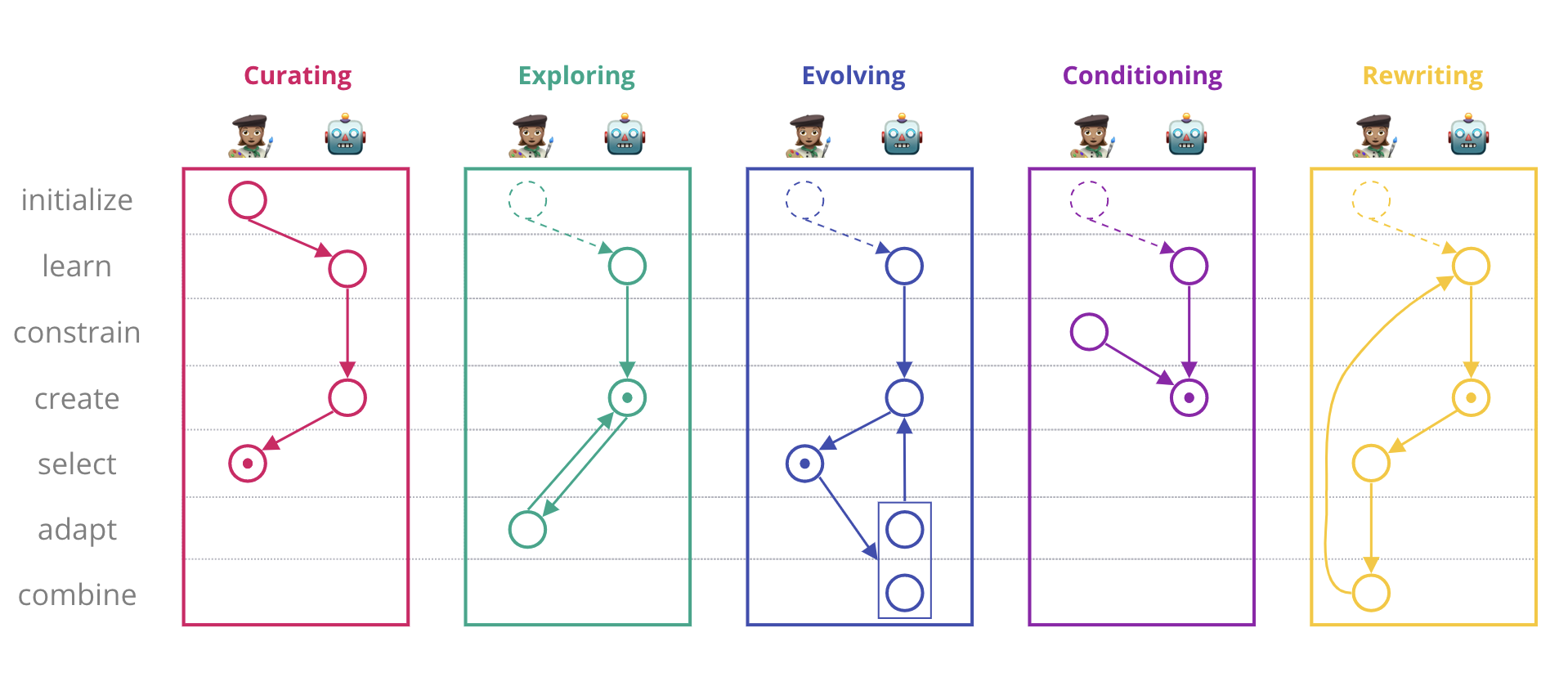}
    \caption{We present the four primary interaction patterns \textit{Curating}, \textit{Exploring}, \textit{Evolving}, and \textit{Conditioning} we identified as part of our preliminary framework \cite{grabe_towards_2022}. We suggest a fifth pattern, \textit{Rewriting}, for discussion.}
    \label{fig:my_label}
\end{figure}

The described project is a work-in-progress and follows the goal of creating co-creative GAN applications. As a foundation, we aim to map existing GAN models to analyze their support of and application in (co-)creative scenarios. While reviewing both co-creative and exclusively technical approaches that could potentially be applied in co-creation, we identified patterns in how we currently (co-)create with GANs \cite{grabe_towards_2022}.
To arrive at this classification, we adapted \citeauthor{spoto_library_2017}'s~\cite{spoto_library_2017} and \citeauthor{muller_mixed_2020}'s~\cite{muller_mixed_2020} framework to a minimal set of actions. Grounded in the algorithmic properties of GANs, the adapted framework distilled four interaction patterns between human and GAN-based computational agents (see Fig.~\ref{fig:my_label}).

The identified patterns help us to make sense of how GANs are applied in creative scenarios, specifically in our chosen application domain of fashion design.
The fashion label Acne Studios' approach of incorporating GAN generations
into their Fall 2020 Menswear Collection falls into the simplest interaction pattern, Curating. The designer(s) chose textures from a GAN's sampled designs without further interacting with the system.
The results received the critique of being a ``math-crunched amalgam of all previous Acne Studio collections.''\footnote{\href{https://www.vogue.com/fashion-shows/fall-2020-menswear/acne-studios}{https://www.vogue.com/fashion-shows/fall-2020-menswear/acne-studios}}
While this points to the characteristic of a GAN's designs lying within the training data distribution, deducting from the example that GANs have nothing novel to add to creative design processes
might be a foregone conclusion.
Rather, we suggest to approach GANs' abilities as a design tool, allowing for the exploration the emerging space of \textit{between} cultural artifacts, elevating the phenotype-genotype mapping to a high-dimensional level for interaction, 
and maybe even offering new forms of a machine-specific creativity.
Utilizing these GAN-specific properties carries the potential of providing novel ways for interactive co-creation.

Differentiated control of GANs applied to fashion design has been achieved through constraining the model with the encoding of a text description \cite{zhu_be_2017}, sketch drawings \cite{zhao_compensation_2018},
or color, texture, and shape inputs through separate losses in the loss function \cite{yildirim_disentangling_2018}. These approaches can all be understood as Conditioning. The human provides the GAN with a constraint in the beginning without further reacting to the model's creation. Hence, no iterative loop of actors replying to each others' creations is repeated.
However, one might postulate that the human could learn the \textit{constraint language} of the GAN by repeating the process. The GAN, however, has neither the option to adapt the design, nor to re-learn. While the former is the case in the Exploring and Evolving pattern, the re-learning during interaction based on human input has been less explored. \citeauthor{bau_rewriting_2020}~\cite{bau_rewriting_2020} present the idea of ``Rewriting'' GANs, suggesting to update GANs' weights based on human's adjustments, which we here map as a possible fifth interaction pattern in Fig.~\ref{fig:my_label}. Applied in design processes, patterns could be interpreted as a form of designer modelling \cite{liapis_can_2016} remembering the user's preferences through learning. 
Together with the other iterative patterns of Exploring end Evolving, which have been practically implemented in small-scale experiments of low-resolution artifacts~\cite{xin_object_2021,bontrager_deep_2018,tejeda_ocampo_appendix_2021} 
mainly with the purpose of probing the algorithm, the research direction suggests the question: 
\textit{How do GANs offer new possibilities for personalized design processes?}
The pattern of Rewriting expands the collaborative process from designing \textit{with} to designing the GAN \textit{itself} during co-creation.

Underlying is the question of how we define the agents interacting in co-creative GAN applications and their creative agency. While we consider supporting algorithms, such as interactive genetic algorithms, as part of the computational agent in the identified patterns~\cite{grabe_towards_2022}, how do we go about generation procedures that include other artificially intelligent components? For example, when a style recognition model steers the generation towards a style identified in an unsupervised manner, would that possibly reveal a machine-specific understanding of style \cite{grabe_fashion_2022}?
How generative deep learning is incorporated in co-creative fashion design matters, as clothing stands in an intimate relation to human beings, conveying meaning beyond its material properties using its own language, that the human eye is trained to read~\cite{hollander_seeing_1993}. Hence, research is required to investigate how humans make sense of generative models that participate in its design and to reflect on co-creative outcomes of human-GAN interaction.

The goal of the project is to develop GANs for the co-creation between fashion designers and machines. Through user studies, we plan to determine the requirements for conscious and intentional co-creation. We are aware that many datasets available for training generative fashion design models consist of social media and catalog images, highly biased to the presented populations. A crucial part is therefore to enable designers to discuss the data’s effect on how stability of a system is reached with regards to design diversity, such as achieving desirable silhouettes while still considering diverse body forms~\cite{hsiao_vibe_2019}.

Fashion designers’ expertise can inform the development of applications, that might also become relevant for consumers or other domains. By better understanding how GANs function, and looking at their underlying features, designers may ask how human perception relates to non-human perception and how fashion could look otherwise. Investigating human-machine collaboration in the creative design process contributes to the contemporary debate about creativity and authorship~\cite{epstein_who_2020}.
The proposed research applies a holistic view to the practical issue of applying generative deep learning in the creative field. By bridging the designer’s needs and technological capabilities, it aims to develop accountable technology, challenging the status quo of the development of creative AI systems.



\bibliographystyle{ACM-Reference-Format}
\bibliography{references.bib}


\begin{thebibliography}{34}


\ifx \showCODEN    \undefined \def \showCODEN     #1{\unskip}     \fi
\ifx \showDOI      \undefined \def \showDOI       #1{#1}\fi
\ifx \showISBNx    \undefined \def \showISBNx     #1{\unskip}     \fi
\ifx \showISBNxiii \undefined \def \showISBNxiii  #1{\unskip}     \fi
\ifx \showISSN     \undefined \def \showISSN      #1{\unskip}     \fi
\ifx \showLCCN     \undefined \def \showLCCN      #1{\unskip}     \fi
\ifx \shownote     \undefined \def \shownote      #1{#1}          \fi
\ifx \showarticletitle \undefined \def \showarticletitle #1{#1}   \fi
\ifx \showURL      \undefined \def \showURL       {\relax}        \fi
\providecommand\bibfield[2]{#2}
\providecommand\bibinfo[2]{#2}
\providecommand\natexlab[1]{#1}
\providecommand\showeprint[2][]{arXiv:#2}

\bibitem[Bau et~al\mbox{.}(2020)]%
        {bau_rewriting_2020}
\bibfield{author}{\bibinfo{person}{David Bau}, \bibinfo{person}{Steven Liu},
  \bibinfo{person}{Tongzhou Wang}, \bibinfo{person}{Jun-Yan Zhu}, {and}
  \bibinfo{person}{Antonio Torralba}.} \bibinfo{year}{2020}\natexlab{}.
\newblock \showarticletitle{Rewriting a {Deep} {Generative} {Model}}. In
  \bibinfo{booktitle}{\emph{Proceedings of the {European} {Conference} on
  {Computer} {Vision} ({ECCV})}}.
\newblock


\bibitem[Benjamin et~al\mbox{.}(2021)]%
        {benjamin_machine_2021}
\bibfield{author}{\bibinfo{person}{Jesse~Josua Benjamin}, \bibinfo{person}{Arne
  Berger}, \bibinfo{person}{Nick Merrill}, {and} \bibinfo{person}{James
  Pierce}.} \bibinfo{year}{2021}\natexlab{}.
\newblock \showarticletitle{Machine learning uncertainty as a design material:
  {A} post-phenomenological inquiry}. \bibinfo{publisher}{Association for
  Computing Machinery}.
\newblock
\showISBNx{978-1-4503-8096-6}
\urldef\tempurl%
\url{https://doi.org/10.1145/3411764.3445481}
\showDOI{\tempurl}


\bibitem[Bontrager et~al\mbox{.}(2018a)]%
        {bontrager_deep_2018}
\bibfield{author}{\bibinfo{person}{Philip Bontrager}, \bibinfo{person}{Wending
  Lin}, \bibinfo{person}{Julian Togelius}, {and} \bibinfo{person}{Sebastian
  Risi}.} \bibinfo{year}{2018}\natexlab{a}.
\newblock \showarticletitle{Deep {Interactive} {Evolution}}. In
  \bibinfo{booktitle}{\emph{International {Conference} on {Computational}
  {Intelligence} in {Music}, {Sound}, {Art} and {Design} : {EvoMUSART} 2018}}.
\newblock


\bibitem[Bontrager et~al\mbox{.}(2018b)]%
        {bontrager_deepmasterprints_2018}
\bibfield{author}{\bibinfo{person}{Philip Bontrager}, \bibinfo{person}{Aditi
  Roy}, \bibinfo{person}{Julian Togelius}, \bibinfo{person}{Nasir Memon}, {and}
  \bibinfo{person}{Arun Ross}.} \bibinfo{year}{2018}\natexlab{b}.
\newblock \showarticletitle{{DeepMasterPrints}: {Generating} {MasterPrints} for
  {Dictionary} {Attacks} via {Latent} {Variable} {Evolution}*}. In
  \bibinfo{booktitle}{\emph{2018 {IEEE} 9th {International} {Conference} on
  {Biometrics} {Theory}, {Applications} and {Systems} ({BTAS})}}.
  \bibinfo{pages}{1--9}.
\newblock
\urldef\tempurl%
\url{https://doi.org/10.1109/BTAS.2018.8698539}
\showDOI{\tempurl}
\newblock
\shownote{ISSN: 2474-9699}.


\bibitem[Buschek et~al\mbox{.}(2021)]%
        {buschek_nine_2021}
\bibfield{author}{\bibinfo{person}{Daniel Buschek}, \bibinfo{person}{Lukas
  Mecke}, \bibinfo{person}{Florian Lehmann}, {and} \bibinfo{person}{Hai Dang}.}
  \bibinfo{year}{2021}\natexlab{}.
\newblock \showarticletitle{Nine {Potential} {Pitfalls} when {Designing}
  {Human}-{AI} {Co}-{Creative} {Systems}}. In \bibinfo{booktitle}{\emph{Joint
  {Proceedings} of the {ACM} {IUI} 2021 {Workshops}}}.
  \bibinfo{address}{College Station, USA}.
\newblock
\urldef\tempurl%
\url{http://arxiv.org/abs/2104.00358}
\showURL{%
\tempurl}


\bibitem[Denton et~al\mbox{.}(2019)]%
        {denton_image_2019}
\bibfield{author}{\bibinfo{person}{Emily Denton}, \bibinfo{person}{Ben
  Hutchinson}, \bibinfo{person}{Margaret Mitchell}, \bibinfo{person}{Timnit
  Gebru}, {and} \bibinfo{person}{Andrew Zaldivar}.}
  \bibinfo{year}{2019}\natexlab{}.
\newblock \showarticletitle{Image {Counterfactual} {Sensitivity} {Analysis} for
  {Detecting} {Unintended} {Bias}}.
\newblock  (\bibinfo{date}{June} \bibinfo{year}{2019}).
\newblock
\urldef\tempurl%
\url{http://arxiv.org/abs/1906.06439}
\showURL{%
\tempurl}


\bibitem[Deterding et~al\mbox{.}(2017)]%
        {deterding_mixed-initiative_2017}
\bibfield{author}{\bibinfo{person}{Sebastian Deterding},
  \bibinfo{person}{Jonathan Hook}, \bibinfo{person}{Rebecca Fiebrink},
  \bibinfo{person}{Marco Gillies}, \bibinfo{person}{Jeremy Gow},
  \bibinfo{person}{Memo Akten}, \bibinfo{person}{Gillian Smith},
  \bibinfo{person}{Antonios Liapis}, {and} \bibinfo{person}{Kate Compton}.}
  \bibinfo{year}{2017}\natexlab{}.
\newblock \showarticletitle{Mixed-{Initiative} {Creative} {Interfaces}}. In
  \bibinfo{booktitle}{\emph{Proceedings of the 2017 {CHI} {Conference}
  {Extended} {Abstracts} on {Human} {Factors} in {Computing} {Systems}}}.
  \bibinfo{publisher}{ACM}, \bibinfo{address}{Denver Colorado USA},
  \bibinfo{pages}{628--635}.
\newblock
\showISBNx{978-1-4503-4656-6}
\urldef\tempurl%
\url{https://doi.org/10.1145/3027063.3027072}
\showDOI{\tempurl}


\bibitem[Elgammal et~al\mbox{.}(2017)]%
        {elgammal_can_2017}
\bibfield{author}{\bibinfo{person}{Ahmed Elgammal}, \bibinfo{person}{Bingchen
  Liu}, \bibinfo{person}{Mohamed Elhoseiny}, {and} \bibinfo{person}{Marian
  Mazzone}.} \bibinfo{year}{2017}\natexlab{}.
\newblock \showarticletitle{{CAN}: {Creative} {Adversarial} {Networks},
  {Generating} "{Art}" by {Learning} {About} {Styles} and {Deviating} from
  {Style} {Norms}}.
\newblock  (\bibinfo{date}{June} \bibinfo{year}{2017}).
\newblock
\urldef\tempurl%
\url{http://arxiv.org/abs/1706.07068}
\showURL{%
\tempurl}


\bibitem[Epstein et~al\mbox{.}(2020)]%
        {epstein_who_2020}
\bibfield{author}{\bibinfo{person}{Ziv Epstein}, \bibinfo{person}{Sydney
  Levine}, \bibinfo{person}{David~G. Rand}, {and} \bibinfo{person}{Iyad
  Rahwan}.} \bibinfo{year}{2020}\natexlab{}.
\newblock \showarticletitle{Who {Gets} {Credit} for {AI}-{Generated} {Art}?}
\newblock \bibinfo{journal}{\emph{iScience}} \bibinfo{volume}{23},
  \bibinfo{number}{9} (\bibinfo{date}{Aug.} \bibinfo{year}{2020}),
  \bibinfo{pages}{101515}.
\newblock
\showISSN{2589-0042}
\urldef\tempurl%
\url{https://doi.org/10.1016/j.isci.2020.101515}
\showDOI{\tempurl}


\bibitem[Gajdacz et~al\mbox{.}(2021)]%
        {gajdacz_creablender_2021}
\bibfield{author}{\bibinfo{person}{Miroslav Gajdacz}, \bibinfo{person}{Janet
  Rafner}, \bibinfo{person}{Steven Langsford}, \bibinfo{person}{Arthur Hjorth},
  \bibinfo{person}{Carsten Bergenholtz}, \bibinfo{person}{Michael~Mose
  Biskjaer}, \bibinfo{person}{Lior Noy}, \bibinfo{person}{Sebastian Risi},
  {and} \bibinfo{person}{Jacob Sherson}.} \bibinfo{year}{2021}\natexlab{}.
\newblock \showarticletitle{{CREA}.blender: a {GAN} based casual creator for
  creativity assessment}. In \bibinfo{booktitle}{\emph{Proceedings of the
  {International} {Conference} on {Computational} {Creativity} ({ICCC}
  ’21)}}. \bibinfo{pages}{5}.
\newblock


\bibitem[Goodfellow et~al\mbox{.}(2014)]%
        {goodfellow_generative_2014}
\bibfield{author}{\bibinfo{person}{Ian~J Goodfellow}, \bibinfo{person}{Jean
  Pouget-Abadie}, \bibinfo{person}{Mehdi Mirza}, \bibinfo{person}{Bing Xu},
  \bibinfo{person}{David Warde-Farley}, \bibinfo{person}{Sherjil Ozair},
  \bibinfo{person}{Aaron Courville}, {and} \bibinfo{person}{Yoshua Bengio}.}
  \bibinfo{year}{2014}\natexlab{}.
\newblock \showarticletitle{Generative {Adversarial} {Nets}}. In
  \bibinfo{booktitle}{\emph{{NIPS}'14: {Proceedings} of the 27th
  {International} {Conference} on {Neural} {Information} {Processing}
  {Systems}}}, Vol.~\bibinfo{volume}{2}. \bibinfo{pages}{2672--2680}.
\newblock
\urldef\tempurl%
\url{http://www.github.com/goodfeli/adversarial}
\showURL{%
\tempurl}


\bibitem[Grabe et~al\mbox{.}(2022a)]%
        {grabe_towards_2022}
\bibfield{author}{\bibinfo{person}{Imke Grabe}, \bibinfo{person}{Miguel
  González-Duque}, \bibinfo{person}{Sebastian Risi}, {and}
  \bibinfo{person}{Jichen Zhu}.} \bibinfo{year}{2022}\natexlab{a}.
\newblock \showarticletitle{Towards a {Framework} for {Human}-{AI}
  {Interaction} {Patterns} in {Co}-{Creative} {GAN} {Applications}}. In
  \bibinfo{booktitle}{\emph{Joint {Proceedings} of the {ACM} {IUI} {Workshops}
  2022, {March} 2022, {Helsinki}, {Finland}}}. \bibinfo{pages}{11}.
\newblock
\urldef\tempurl%
\url{https://hai-gen.github.io/2022/papers/paper-HAIGEN-GrabeImke.pdf}
\showURL{%
\tempurl}


\bibitem[Grabe et~al\mbox{.}(2022b)]%
        {grabe_fashion_2022}
\bibfield{author}{\bibinfo{person}{Imke Grabe}, \bibinfo{person}{Jichen Zhu},
  {and} \bibinfo{person}{Manex Agirrezabal}.} \bibinfo{year}{2022}\natexlab{b}.
\newblock \showarticletitle{Fashion {Style} {Generation}: {Evolutionary}
  {Search} with {Gaussian} {Mixture} {Models} in the {Latent} {Space}}. In
  \bibinfo{booktitle}{\emph{International {Conference} on {Computational}
  {Intelligence} in {Music}, {Sound}, {Art} and {Design} : {EvoMUSART} 2022}}.
  \bibinfo{pages}{16}.
\newblock


\bibitem[Hollander(1993)]%
        {hollander_seeing_1993}
\bibfield{author}{\bibinfo{person}{Anne Hollander}.}
  \bibinfo{year}{1993}\natexlab{}.
\newblock \bibinfo{booktitle}{\emph{Seeing {Through} {Clothes}}}.
\newblock \bibinfo{publisher}{University of California Press}.
\newblock
\showISBNx{978-0-520-08231-1}
\newblock
\shownote{Google-Books-ID: T7cwDwAAQBAJ}.


\bibitem[Hsiao and Grauman(2019)]%
        {hsiao_vibe_2019}
\bibfield{author}{\bibinfo{person}{Wei-Lin Hsiao} {and}
  \bibinfo{person}{Kristen Grauman}.} \bibinfo{year}{2019}\natexlab{}.
\newblock \showarticletitle{{ViBE}: {Dressing} for {Diverse} {Body} {Shapes}}.
\newblock  (\bibinfo{date}{Dec.} \bibinfo{year}{2019}).
\newblock
\urldef\tempurl%
\url{http://arxiv.org/abs/1912.06697}
\showURL{%
\tempurl}


\bibitem[Hughes et~al\mbox{.}(2021)]%
        {hughes_generative_2021}
\bibfield{author}{\bibinfo{person}{Rowan~T. Hughes}, \bibinfo{person}{Liming
  Zhu}, {and} \bibinfo{person}{Tomasz Bednarz}.}
  \bibinfo{year}{2021}\natexlab{}.
\newblock \showarticletitle{Generative {Adversarial} {Networks}–{Enabled}
  {Human}–{Artificial} {Intelligence} {Collaborative} {Applications} for
  {Creative} and {Design} {Industries}: {A} {Systematic} {Review} of {Current}
  {Approaches} and {Trends}}.
\newblock \bibinfo{journal}{\emph{Frontiers in Artificial Intelligence}}
  \bibinfo{volume}{4} (\bibinfo{date}{April} \bibinfo{year}{2021}),
  \bibinfo{pages}{604234}.
\newblock
\showISSN{2624-8212}
\urldef\tempurl%
\url{https://doi.org/10.3389/frai.2021.604234}
\showDOI{\tempurl}


\bibitem[Karras et~al\mbox{.}(2019)]%
        {karras_style-based_2019}
\bibfield{author}{\bibinfo{person}{Tero Karras}, \bibinfo{person}{Samuli
  Laine}, {and} \bibinfo{person}{Timo Aila}.} \bibinfo{year}{2019}\natexlab{}.
\newblock \showarticletitle{A {Style}-{Based} {Generator} {Architecture} for
  {Generative} {Adversarial} {Networks}}. In
  \bibinfo{booktitle}{\emph{Proceedings of the {IEEE}/{CVF} conference on
  computer vision and pattern recognition}}. \bibinfo{pages}{4401--4410}.
\newblock
\urldef\tempurl%
\url{http://arxiv.org/abs/1812.04948}
\showURL{%
\tempurl}


\bibitem[Liapis et~al\mbox{.}(2016)]%
        {liapis_can_2016}
\bibfield{author}{\bibinfo{person}{Antonios Liapis},
  \bibinfo{person}{Georgios~N Yannakakis}, \bibinfo{person}{Constantine
  Alexopoulos}, {and} \bibinfo{person}{Phil Lopes}.}
  \bibinfo{year}{2016}\natexlab{}.
\newblock \showarticletitle{{CAN} {COMPUTERS} {FOSTER} {HUMAN} {USERS}’
  {CREATIVITY}? {THEORY} {AND} {PRAXIS} {OF} {MIXED}- {INITIATIVE}
  {CO}-{CREATIVITY}}.
\newblock  (\bibinfo{year}{2016}), \bibinfo{pages}{17}.
\newblock


\bibitem[Muller et~al\mbox{.}(2020)]%
        {muller_mixed_2020}
\bibfield{author}{\bibinfo{person}{Michael Muller}, \bibinfo{person}{Justin~D
  Weisz}, {and} \bibinfo{person}{Werner Geyer}.}
  \bibinfo{year}{2020}\natexlab{}.
\newblock \showarticletitle{Mixed {Initiative} {Generative} {AI} {Interfaces}:
  {An} {Analytic} {Framework} for {Generative} {AI} {Applications}}. In
  \bibinfo{booktitle}{\emph{Proceedings of the {Workshop} {The} {Future} of
  {Co}-{Creative} {Systems} - {A} {Workshop} on {Human}-{Computer}
  {Co}-{Creativity} of the 11th {International} {Conference} on {Computational}
  {Creativity} ({ICCC} 2020)}}.
\newblock
\urldef\tempurl%
\url{https://computationalcreativity.net/workshops/cocreative-iccc20/papers/Future_of_co-creative_systems_185.pdf}
\showURL{%
\tempurl}


\bibitem[Rostamzadeh et~al\mbox{.}(2018)]%
        {rostamzadeh_fashion-gen_2018}
\bibfield{author}{\bibinfo{person}{Negar Rostamzadeh},
  \bibinfo{person}{Seyedarian Hosseini}, \bibinfo{person}{Thomas Boquet},
  \bibinfo{person}{Wojciech Stokowiec}, \bibinfo{person}{Ying Zhang},
  \bibinfo{person}{Christian Jauvin}, {and} \bibinfo{person}{Chris Pal}.}
  \bibinfo{year}{2018}\natexlab{}.
\newblock \showarticletitle{Fashion-{Gen}: {The} {Generative} {Fashion}
  {Dataset} and {Challenge}}.
\newblock \bibinfo{journal}{\emph{arXiv:1806.08317 [cs, stat]}}
  (\bibinfo{date}{July} \bibinfo{year}{2018}).
\newblock
\urldef\tempurl%
\url{http://arxiv.org/abs/1806.08317}
\showURL{%
\tempurl}
\newblock
\shownote{arXiv: 1806.08317}.


\bibitem[Roy et~al\mbox{.}(2021)]%
        {roy_survey_2021}
\bibfield{author}{\bibinfo{person}{William Roy}, \bibinfo{person}{Glen Kelly},
  \bibinfo{person}{Robert Leer}, {and} \bibinfo{person}{Frederick Ricardo}.}
  \bibinfo{year}{2021}\natexlab{}.
\newblock \showarticletitle{A {Survey} on {Adversarial} {Image} {Synthesis}}.
\newblock \bibinfo{journal}{\emph{ACM Comput. Surv.}} (\bibinfo{date}{July}
  \bibinfo{year}{2021}).
\newblock
\urldef\tempurl%
\url{http://arxiv.org/abs/2106.16056}
\showURL{%
\tempurl}
\newblock
\shownote{arXiv: 2106.16056}.


\bibitem[Schrum et~al\mbox{.}(2020)]%
        {schrum_interactive_2020}
\bibfield{author}{\bibinfo{person}{Jacob Schrum}, \bibinfo{person}{Jake
  Gutierrez}, \bibinfo{person}{Vanessa Volz}, \bibinfo{person}{Jialin Liu},
  \bibinfo{person}{Simon Lucas}, {and} \bibinfo{person}{Sebastian Risi}.}
  \bibinfo{year}{2020}\natexlab{}.
\newblock \showarticletitle{Interactive {Evolution} and {Exploration} {Within}
  {Latent} {Level}-{Design} {Space} of {Generative} {Adversarial} {Networks}}.
  In \bibinfo{booktitle}{\emph{Proceedings of the 2020 {Genetic} and
  {Evolutionary} {Computation} {Conference}}}. \bibinfo{pages}{148--156}.
\newblock
\urldef\tempurl%
\url{https://doi.org/10.1145/3377930.3389821}
\showURL{%
\tempurl}
\newblock
\shownote{arXiv: 2004.00151}.


\bibitem[Shamsolmoali et~al\mbox{.}(2021)]%
        {shamsolmoali_image_2021}
\bibfield{author}{\bibinfo{person}{Pourya Shamsolmoali},
  \bibinfo{person}{Masoumeh Zareapoor}, \bibinfo{person}{Eric Granger},
  \bibinfo{person}{Huiyu Zhou}, \bibinfo{person}{Ruili Wang},
  \bibinfo{person}{M.~Emre Celebi}, {and} \bibinfo{person}{Jie Yang}.}
  \bibinfo{year}{2021}\natexlab{}.
\newblock \showarticletitle{Image synthesis with adversarial networks: {A}
  comprehensive survey and case studies}.
\newblock \bibinfo{journal}{\emph{Information Fusion}}  \bibinfo{volume}{72}
  (\bibinfo{date}{Aug.} \bibinfo{year}{2021}), \bibinfo{pages}{126--146}.
\newblock
\showISSN{1566-2535}
\urldef\tempurl%
\url{https://doi.org/10.1016/j.inffus.2021.02.014}
\showDOI{\tempurl}


\bibitem[Shen et~al\mbox{.}(2020)]%
        {shen_interpreting_2020}
\bibfield{author}{\bibinfo{person}{Yujun Shen}, \bibinfo{person}{Jinjin Gu},
  \bibinfo{person}{Xiaoou Tang}, {and} \bibinfo{person}{Bolei Zhou}.}
  \bibinfo{year}{2020}\natexlab{}.
\newblock \showarticletitle{Interpreting the {Latent} {Space} of {GANs} for
  {Semantic} {Face} {Editing}}. In \bibinfo{booktitle}{\emph{2020 {IEEE}/{CVF}
  {Conference} on {Computer} {Vision} and {Pattern} {Recognition} ({CVPR})}}.
  \bibinfo{publisher}{IEEE}, \bibinfo{address}{Seattle, WA, USA},
  \bibinfo{pages}{9240--9249}.
\newblock
\showISBNx{978-1-72817-168-5}
\urldef\tempurl%
\url{https://doi.org/10.1109/CVPR42600.2020.00926}
\showDOI{\tempurl}


\bibitem[Spoto and Oleynik(2017)]%
        {spoto_library_2017}
\bibfield{author}{\bibinfo{person}{Angie Spoto} {and} \bibinfo{person}{Natalia
  Oleynik}.} \bibinfo{year}{2017}\natexlab{}.
\newblock \bibinfo{title}{Library of {Mixed}-{Initiative} {Creative}
  {Interfaces}}.
\newblock
\newblock
\urldef\tempurl%
\url{http://mici.codingconduct.cc/}
\showURL{%
\tempurl}


\bibitem[Tejeda~Ocampo et~al\mbox{.}(2021)]%
        {tejeda_ocampo_appendix_2021}
\bibfield{author}{\bibinfo{person}{Carlos Tejeda~Ocampo},
  \bibinfo{person}{Armando López-Cuevas}, {and} \bibinfo{person}{Hugo
  Terashima-Marin}.} \bibinfo{year}{2021}\natexlab{}.
\newblock \showarticletitle{Appendix: {Improving} {Deep} {Interactive}
  {Evolution} with {Style}-{Based} {Generator} for {Artistic} {Expression} and
  {Creative} {Exploration}-{Appendix}}.
\newblock  (\bibinfo{year}{2021}).
\newblock
\urldef\tempurl%
\url{https://doi.org/10.3390/exx010005}
\showDOI{\tempurl}


\bibitem[Wu et~al\mbox{.}(2017)]%
        {wu_survey_2017}
\bibfield{author}{\bibinfo{person}{Xian Wu}, \bibinfo{person}{Kun Xu}, {and}
  \bibinfo{person}{Peter Hall}.} \bibinfo{year}{2017}\natexlab{}.
\newblock \showarticletitle{A survey of image synthesis and editing with
  generative adversarial networks}.
\newblock \bibinfo{journal}{\emph{Tsinghua Science and Technology}}
  \bibinfo{volume}{22}, \bibinfo{number}{6} (\bibinfo{date}{Dec.}
  \bibinfo{year}{2017}), \bibinfo{pages}{660--674}.
\newblock
\showISSN{1007-0214}
\urldef\tempurl%
\url{https://doi.org/10.23919/TST.2017.8195348}
\showDOI{\tempurl}
\newblock
\shownote{Conference Name: Tsinghua Science and Technology}.


\bibitem[Xin and Arakawa(2021)]%
        {xin_object_2021}
\bibfield{author}{\bibinfo{person}{Chen Xin} {and} \bibinfo{person}{Kaoru
  Arakawa}.} \bibinfo{year}{2021}\natexlab{}.
\newblock \showarticletitle{Object {Design} {System} by {Interactive}
  {Evolutionary} {Computation} {Using} {GAN} with {Contour} {Images}}. In
  \bibinfo{booktitle}{\emph{Human {Centred} {Intelligent} {Systems} -
  {Proceedings} of {KES}-{HCIS} 2021 {Conference}}},
  Vol.~\bibinfo{volume}{244}. \bibinfo{publisher}{Springer Singapore},
  \bibinfo{address}{Singapore}, \bibinfo{pages}{66--75}.
\newblock
\showISBNx{9789811632631 9789811632648}
\urldef\tempurl%
\url{https://doi.org/10.1007/978-981-16-3264-8_7}
\showDOI{\tempurl}
\newblock
\shownote{Series Title: Smart Innovation, Systems and Technologies}.


\bibitem[Yang et~al\mbox{.}(2020)]%
        {yang_re-examining_2020}
\bibfield{author}{\bibinfo{person}{Qian Yang}, \bibinfo{person}{Aaron
  Steinfeld}, \bibinfo{person}{Carolyn Rosé}, {and} \bibinfo{person}{John
  Zimmerman}.} \bibinfo{year}{2020}\natexlab{}.
\newblock \showarticletitle{Re-examining {Whether}, {Why}, and {How}
  {Human}-{AI} {Interaction} {Is} {Uniquely} {Difficult} to {Design}}. In
  \bibinfo{booktitle}{\emph{Proceedings of the 2020 {CHI} {Conference} on
  {Human} {Factors} in {Computing} {Systems}}} \emph{(\bibinfo{series}{{CHI}
  '20})}. \bibinfo{publisher}{Association for Computing Machinery},
  \bibinfo{address}{New York, NY, USA}, \bibinfo{pages}{1--13}.
\newblock
\showISBNx{978-1-4503-6708-0}
\urldef\tempurl%
\url{https://doi.org/10.1145/3313831.3376301}
\showDOI{\tempurl}


\bibitem[Yildirim et~al\mbox{.}(2018)]%
        {yildirim_disentangling_2018}
\bibfield{author}{\bibinfo{person}{Gökhan Yildirim}, \bibinfo{person}{Calvin
  Seward}, {and} \bibinfo{person}{Urs Bergmann}.}
  \bibinfo{year}{2018}\natexlab{}.
\newblock \showarticletitle{Disentangling {Multiple} {Conditional} {Inputs} in
  {GANs}}. In \bibinfo{booktitle}{\emph{{KDD} 2018 {Conference} {AI} for
  {Fashion} {Workshop}}}.
\newblock
\urldef\tempurl%
\url{http://arxiv.org/abs/1806.07819}
\showURL{%
\tempurl}
\newblock
\shownote{arXiv: 1806.07819}.


\bibitem[Zaltron et~al\mbox{.}(2020)]%
        {zaltron_cg-gan_2020}
\bibfield{author}{\bibinfo{person}{Nicola Zaltron}, \bibinfo{person}{Luisa
  Zurlo}, {and} \bibinfo{person}{Sebastian Risi}.}
  \bibinfo{year}{2020}\natexlab{}.
\newblock \showarticletitle{{CG}-{GAN}: {An} {Interactive} {Evolutionary}
  {GAN}-{Based} {Approach} for {Facial} {Composite} {Generation}}.
\newblock \bibinfo{journal}{\emph{Proceedings of the AAAI Conference on
  Artificial Intelligence}} \bibinfo{volume}{34}, \bibinfo{number}{03}
  (\bibinfo{date}{April} \bibinfo{year}{2020}), \bibinfo{pages}{2544--2551}.
\newblock
\showISSN{2374-3468, 2159-5399}
\urldef\tempurl%
\url{https://doi.org/10.1609/aaai.v34i03.5637}
\showDOI{\tempurl}


\bibitem[Zhang and Zhu(2018)]%
        {zhang_visual_2018}
\bibfield{author}{\bibinfo{person}{Quanshi Zhang} {and}
  \bibinfo{person}{Song-Chun Zhu}.} \bibinfo{year}{2018}\natexlab{}.
\newblock \showarticletitle{Visual {Interpretability} for {Deep} {Learning}: a
  {Survey}}.
\newblock \bibinfo{journal}{\emph{Frontiers of Information Technology \&
  Electronic Engineering volume}} (\bibinfo{date}{Feb.} \bibinfo{year}{2018}).
\newblock
\urldef\tempurl%
\url{http://arxiv.org/abs/1802.00614}
\showURL{%
\tempurl}
\newblock
\shownote{arXiv: 1802.00614}.


\bibitem[Zhao and Ma(2018)]%
        {zhao_compensation_2018}
\bibfield{author}{\bibinfo{person}{Zhenjie Zhao} {and}
  \bibinfo{person}{Xiaojuan Ma}.} \bibinfo{year}{2018}\natexlab{}.
\newblock \showarticletitle{A {Compensation} {Method} of {Two}-{Stage} {Image}
  {Generation} for {Human}-{AI} {Collaborated} {In}-{Situ} {Fashion} {Design}
  in {Augmented} {Reality} {Environment}}. In \bibinfo{booktitle}{\emph{2018
  {IEEE} {International} {Conference} on {Artificial} {Intelligence} and
  {Virtual} {Reality} ({AIVR})}}. \bibinfo{publisher}{IEEE},
  \bibinfo{address}{Taichung, Taiwan}, \bibinfo{pages}{76--83}.
\newblock
\showISBNx{978-1-5386-9269-1}
\urldef\tempurl%
\url{https://doi.org/10.1109/AIVR.2018.00018}
\showDOI{\tempurl}


\bibitem[Zhu et~al\mbox{.}(2017)]%
        {zhu_be_2017}
\bibfield{author}{\bibinfo{person}{Shizhan Zhu}, \bibinfo{person}{Sanja
  Fidler}, \bibinfo{person}{Raquel Urtasun}, \bibinfo{person}{Dahua Lin}, {and}
  \bibinfo{person}{Chen~Change Loy}.} \bibinfo{year}{2017}\natexlab{}.
\newblock \showarticletitle{Be {Your} {Own} {Prada}: {Fashion} {Synthesis} with
  {Structural} {Coherence}}. In \bibinfo{booktitle}{\emph{Proceedings of the
  {IEEE} {International} {Conference} on {Computer} {Vision} ({ICCV})}}.
  \bibinfo{pages}{1680--1688}.
\newblock
\urldef\tempurl%
\url{http://arxiv.org/abs/1710.07346}
\showURL{%
\tempurl}
\newblock
\shownote{arXiv: 1710.07346}.


\end{thebibliography}


\end{document}